   \documentclass[reprint,aps]{revtex4-1}
                   \usepackage{dcolumn}
\usepackage{amsmath}
\usepackage{graphicx}
\usepackage{bm}
\usepackage{color}
\usepackage{comment}
\usepackage{scalefnt}

\newcommand{\beq}{\begin{equation}}
\newcommand{\enq}{\end{equation}}
\newcommand{\beqa}{\begin{eqnarray}}
\newcommand{\beqast}{\begin{eqnarray*}}
\newcommand{\enqa}{\end{eqnarray}}
\newcommand{\enqast}{\end{eqnarray*}}

\providecommand{\U}[1]{\protect\rule{.1in}{.1in}}

\begin{document}

\title{  Analyticity and scaling property of pp and p\=p forward scattering amplitudes}

\author{A. K. Kohara     }

\begin{abstract}
We analyse the pp elastic scattering amplitudes using the recent LHC data, revisiting the model proposed by A. Martin based on analytic continuation and crossing symmetry.  Introducing a new form for the scaling function we show that the data are consistent with the crossing symmetry of the scattering amplitudes at $\sqrt{s}= $ 7 and 8 TeV. The complex amplitude automatically obeys the constraints of dispersion relations and their derivatives. The real part reproduces the zero predicted by A. Martin, which is crucial to describe with precision the differential cross section  in the forward direction at LHC energies. Since the free parameters of the model are energy independent, the analytical form of the amplitude leads to predictions  for higher and asymptotic energies.
\end{abstract}

\affiliation{  {\em Instituto de F\'{\i}sica, Universidade Federal Fluminense \\
 Rio de Janeiro 24210-346, RJ, Brazil }    }

\maketitle
Theoretical and phenomenological approaches for the description of  pp and p\=p elastic scattering aim to determine the dynamics and kinematical dependence of the amplitudes,  described in terms of the two variables $s$ and $t$. 
In Regge theory the rise of the hadronic total cross section is due to  the Pomeron trajectory  linear in $t$, with power dependence on $s$. However for high energies the growth of the total cross section guided by the Froissart bound \cite{Froissart} and by the behaviour of the observed data can be parametrized as a quadratic form in $\log (s)$ such as $\sigma \sim \log^2 (s)$. The form of the differential cross section depends on specific assumptions  for  the real and imaginary  amplitudes, controled by dispersion relations (DR).

In another treatment for the very forward region the scattering amplitude $T^N(s,t)$ is suggested to follow a scaling dependence \cite{Deus}, with $T^N(s,t)/T^N(s,0)=f(\tau)$ where $\tau$ is a combination of $s$ and $t$ variables. Using the $\log^2(s)$ dependence combined with the scaling function $f(\tau)$ the scattering amplitude is then written with the form,
\begin{eqnarray}
T^N(s,t)\sim  iC s \log^2(s) f(\tau)~,
\label{amplitude_pomeron_2}
\end{eqnarray}
with $f(\tau)$ normalized such that $f(0)=1$.  The bounds and constraints  of $f(\tau)$ were formally studied long ago \cite{Deus} in the context of axiomatic field theory, giving $f(\tau)\leq \kappa \exp(\sqrt{|\tau|})$, where $\kappa$ is constant and  the scaling variable is $\tau = t\log^2 s$.
The cross section corresponding to Eq.(\ref{amplitude_pomeron_2}) is not invariant under the transformation, $s \rightarrow -s$ (crossing symmetry).
  According to A. Martin \cite{Martin_Real}, in order to define a complex crossing symmetric function,  Eq.(\ref{amplitude_pomeron_2}) can be modified to 
\begin{eqnarray}
T^N(s,t)\sim  iC s\Big(\log (s)-i\frac{\pi}{2}\Big)^2 f(\tau')~,
\label{amplitude_pomeron_3}
\end{eqnarray}
and the scaling variable is 
$\tau'=t(\log (s)-i\pi/2)^2$.

 However it is not obvious that at LHC energies these amplitudes with crossing are well satisfied.
To test, we propose a generalization of Eq.(\ref{amplitude_pomeron_3}) writing, 
  \begin{eqnarray}
T^N(s,t)=  iC s\Big(\log (s)-i\beta\Big)^2 f(\tau')~,
\label{amplitude_pomeron_4}
\end{eqnarray}
where $\beta$ is a free parameter and
the complex scaling variable is now
\begin{equation}
\tau'=\Big(\log(s)-i\beta\Big)^2t ~,
\label{tauL}
\end{equation}
and we assume the scaling function
\begin{eqnarray}
f(\tau')\equiv e^{\alpha \tau'}=e^{\alpha[\log^2(s)-\beta^2]t-2i\alpha\beta\log (s) t}~,
\label{damping_f}
\end{eqnarray}
where $\alpha$ is a real positive quantity responsible for the shrinkage of the differential cross section, 
and  $f(\tau')$ is a holomorphic function.

Since we study the very forward region where $-t$ goes from 0 to  $-t \sim 2/(\alpha\log^2 s)$, Eq.(\ref{damping_f}) can be approximated as
\begin{equation}
f(\tau')\simeq e^{\alpha[\log^2(s)-\beta^2]t}\Big(1-2i\alpha\beta\log (s)~   t \Big)~.
\label{damping_f1}
\end{equation}
Using Eq.(\ref{damping_f1}) in Eq.(\ref{amplitude_pomeron_4}), the real and imaginary parts are respectively
\begin{eqnarray}
&&T_R(s,t)\simeq C s\log^2 (s)2\beta \times \nonumber \\
&&\Big(\frac{1}{\log (s)}+\alpha\frac{\log^2 (s)-\beta^2}{\log(s)}t\Big)e^{\alpha[\log^2(s) -\beta^2]t}~
\label{TR1}
\end{eqnarray}
and
\begin{eqnarray}
&&T_I(s,t)\simeq  C s\log^2 (s) \times \nonumber \\
&&\Big(\frac{\log^2(s)-\beta^2}{\log^2(s)}-4\alpha\beta^2~t\Big) e^{\alpha[\log^2(s)-\beta^2]t}~.
\label{TI1}
\end{eqnarray}
The term linear on $t$ in the real part accounts for  a zero in the forward range, which corresponds to Martin's zero \cite{Martin}, while in the imaginary amplitude a zero is located in a non physical region, since $\alpha\beta^2$ is positive. 
We can neglect $\beta^2$ terms in the polynomials on $t$ 
 if we assume $\log^2(s)>>\beta^2$, and Eqs.(\ref{TR1}) and (\ref{TI1}) are simplified to
\begin{eqnarray}
T_R(s,t)\simeq C s\log^2 (s) 2\beta\Big(\frac{1}{\log (s)}+\alpha\log (s)~ t\Big)e^{\alpha[\log^2(s)-\beta^2]t}~ \nonumber \\
\label{TR2}
\end{eqnarray}
and
\begin{eqnarray}
T_I(s,t)\simeq  C s\log^2 (s) e^{\alpha[\log^2(s)-5\beta^2] t}~.
\label{TI2}
\end{eqnarray}
The forward amplitudes are written in terms of the quantities $C$, $\alpha$ and $\beta$. The total cross section is given by the optical theorem
\begin{equation}
\sigma = \frac{T_I(s,0)}{s}= C \log^2 s~,
\label{sigma}
\end{equation}
the ratio of the amplitudes at $-t=0$ is
\begin{equation}
\rho = \frac{T_R(s,0)}{T_I(s,0)} =\frac{2\beta}{\log s}~,
\label{rho}
\end{equation}
and the derivatives of the logarithm of the real and imaginary amplitudes at $|t|=0$  are respectively
\begin{equation}
\frac{\partial}{\partial t}\log T_R(s,t)\Big|_{|t|=0} =  \alpha\Big(2\log^2 (s)-\beta^2\Big)\equiv \frac{B_R^{\rm eff}}{2}~,
\label{BR}
\end{equation}
and
\begin{equation}
\frac{\partial}{\partial t}\log T_I(s,t)\Big|_{|t|=0} = \alpha\Big(\log^2 (s)-5\beta^2\Big)\equiv \frac{B_I^{\rm eff}}{2}~,
\label{BI}
\end{equation}
where $B_R^{\rm eff}$ and $B_I^{\rm eff}$ are refereed to as the effective slopes of the real and imaginary  amplitudes.
These derivatives determine the average slope of the differential cross section at $|t|=0$, which in terms of Eqs. (\ref{rho}), (\ref{BR}) and (\ref{BI}) is given by
\begin{eqnarray}
B\equiv\frac{d}{d t}\log\Big(\frac{d\sigma}{dt}\Big)\Big|_{|t|=0} =   \frac{1}{\rho^2+1}\Big(\rho^2B_R^{\rm eff}+B_I^{\rm eff}\Big)~. 
\label{slope}
\end{eqnarray}

Assuming the pp and p\=p total cross section as equal, the application of  DR  is automatically satisfied,  and since $\sigma =C \log^2(s)$, $\rho$ is given by $\rho =\pi/\log(s)$ \cite{Kinoshita}, which compared with Eq.(\ref{rho}) automatically imposes $\beta=\pi/2$.

Since the derivative of the imaginary part is known by Eq.(\ref{BI}), the application of DR  for the derivatives of the amplitudes \cite{EF2007} can be performed, while the crossing symmetry implies that only the  derivative of the even combination is non zero. We obtain
\begin{eqnarray}
&&\frac{\partial T_R^{\rm (DR)}(s,t)}{\partial t}\Big|_0 = \frac{  s^2}{\pi}{\bf P}\int_{2m^2}^{\infty} \frac{ 2}{s'(s'^2-s^2)}\frac{\partial T_I(s',t)}{\partial t}\Big|_0 \nonumber \\
&&=\frac{2s^2}{\pi} \alpha C ~{\bf P}\int_{2m^2}^{\infty} \frac{\log^2 (s')}{s'^2-s^2} \Big(\log^2(s')-\frac{5}{4}\pi^2\Big)~.
\label{DDR_SLOPES}
\end{eqnarray}
 The principal value integrals can be calculated exactly \cite{DDR_EXACT} and the result for  Eq.(\ref{DDR_SLOPES}) divided by $T_R(s,0)$ is
\begin{eqnarray}
&&\frac{\partial }{\partial t}\log T_R^{\rm (DR)}(s,t)\Big|_{|t|=0} = \alpha~\Big[2 \log^2(s)-\frac{\pi^2}{4} \Big]~, \nonumber \\
\label{DDR_SLOPES_1}
\end{eqnarray}
that compared with Eq.(\ref{BR}) confirms $\beta=\pi/2$.

Before we move to the phenomenology of the amplitudes we comment on the differences between the complete form Eq.(\ref{damping_f})  and the approximated form Eq. (\ref{damping_f1}). The former has trigonometric functions with more involved intrinsic $s$ and $t$ dependences which is not realistic for the analysed $t$-range, while the latter, expanded to first order, is much simpler. As we show below  the data seems to accept the approximate form. 

To obtain the quantities  $C$, $\alpha$ and $\beta$  for Eqs.(\ref{TR2}) and (\ref{TI2}) we fit the experimental data at $\sqrt{s}=$ 8 and 7 TeV from Atlas \cite{A7,A8} and Totem \cite{T7,T8} Collaborations (four datasets) using 
\begin{eqnarray}   
\frac{d\sigma}{dt}=\frac{1}{16\pi s^2} |T(s,t)|^2  ~   ,
\end{eqnarray}
where the amplitude $T(s,t)$ contains  nuclear and Coulomb interactions. 

The datasets have been  analysed in limited $t$-range ($0< |t|< 0.2$) GeV$^2$ chosen differently for each of them in order to account for the stability of  $\beta$, which in our analysis is manifestly positive. The positiveness of the real amplitude for $|t|$ near zero was recently proved by A. Martin and T. T. Wu \cite{Martin_2017}, and this confirms our results for $\beta$.

We obtain that the four datasets analysed are well represented by Eqs.(\ref{TR2}) and (\ref{TI2}) with small $\chi^2/ndf$.  The fits show that  $\beta$ is compatible with $\pi/2$ for all datasets. This parameter is related with the phase of the complex nuclear amplitude. Thus, if we consider $\beta=\pi/2$ as an input in Eq.(\ref{amplitude_pomeron_4}) the scattering amplitude becomes crossing symmetric under $s\to u$ for fixed $-t$. 

We obtain a common value $\alpha \simeq 0.031$ GeV$^{-2}$ for the four  datasets. This parameter seems to be constant under energy variation, supporting the idea of the scaling function $f(\tau')$. On the other hand the normalization parameter $C$ presents small deviations (up to $5\%$) among different experimental sets. However, for each experimental group for different energies, the  deviations in $C$ are smaller. 

Thus $C$ may be considered  a constant quantity and in order to predict values for higher energies we take an average of  the values of the four datasets, obtaining $C=0.3072$ mb  and $\alpha=0.0316$ GeV$^{-2}$.   In Table \ref{predictions} we show the predicted values for $\sqrt{s}=$ 13, 14 and 57 TeV.

\begin{figure}
\includegraphics[scale=0.35]{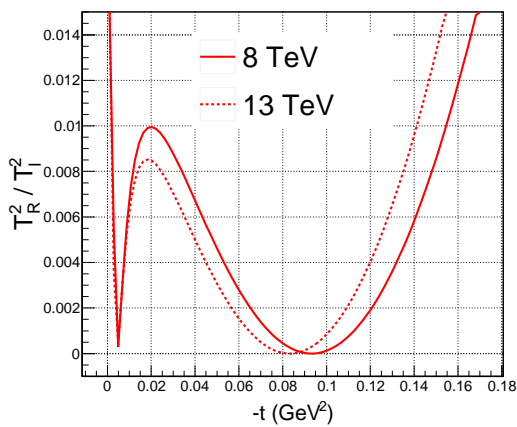}
\caption{The figure represents $T_R^2/T_I^2$. The narrow peak near $|t|=0$ is due to the Coulomb amplitude for pp. The solid curve fits the 8 TeV data very accurately, and the dashed curve is our prediction for 13 TeV. According to Eq.(\ref{tR0})  the position of Martin's zero  moves towards the origin as the energy increases.}
\label{TR2_TI2}
\end{figure}

An important feature of the model is Martin's zero in the real part \cite{Martin},  given by Eq.(\ref{TR2}), with
\begin{eqnarray}
|t_R|=\frac{1}{\alpha\log^2 s}~.
\label{tR0}
\end{eqnarray}
In Fig.\ref{TR2_TI2} we show the ratio $T_R^2/T_I^2$ for $\sqrt{s}=$ 8 and 13 TeV. We observe that the concave structure is created by the zero of the real part and the magnitude of the real slope.    Our treatment is limited to the very forward region, and since for larger values of $-t$ other corrective terms in the amplitudes may play important roles, the position of the zero could be slightly changed.

\begin{table}[t]
\scalefont{1.0}
\begin{center}
 \vspace{0.5cm}
\begin{tabular}{  c c c c c c c}
\hline
\hline
        &           &  &  &   &        & \\
\hline 
  $\sqrt{s}$     &  $\sigma$ & $\rho$& $ B_I^{\rm eff}$  & $ B_R^{\rm eff}$  & $t_R$     \\
     (GeV)           &   (mb) &   & ( GeV$^{-2})$ & ( GeV$^{-2})$& ( GeV$^{2})$    
 \\
 \hline
   13                              &  110.3  & 0.166     & 21.91 & 45.21  & 0.088 & \\
\hline
   14                     & 112.00    &  0.166     &  22.26 &  45.93  & 0.087 &\\
 \hline
    57                    & 147.36    &  0.143     &  29.54 & 60.48  &
0.066&   \\
 \hline

 \end{tabular}
 \caption{ Predictions of forward quantities for $\sqrt{s}=$ 13, 14 and 57 TeV. The quantities $\beta = \pi/2$, $\alpha=0.0316$ GeV$^{-2}$ and $C=0.3072$ mb are fixed. }  
\label{predictions}
\end{center}
\end{table}

The simplicity of the amplitude makes it applicable to Glauber formalism for p-air collisions in cosmic ray experiments.

The asymptotic behaviour of the parameters is crucial to understand whether the proton eventually behaves as a black or a gray disk. For asymptotic energies we obtain the ratio 
\begin{equation}
\frac{\sigma}{(\hbar c)^216\pi B_I^{\rm eff}}\to \frac{C}{(\hbar c)^2 32\pi\alpha}\simeq \frac{1}{4}~,
\label{ratio}
\end{equation}
which is equivalent to  $\sigma_{\rm el}/\sigma\simeq 1/4$~. Thus in the present model the proton behaves as a grey disk, and the inelastic processes correspond to 3/4 of the total. Also, the ratio $B_R^{\rm eff}/B_I^{\rm eff} \to 2$ predicts stronger slope for the real part. A systematic analysis of the asymptotic regime  \cite{Menon} indicates a favorite scenario  compatible with a gray disk with the ratio $\sigma_{\rm el}/\sigma=0.30\pm 0.12$.

The difficulties in the determination of the parameter $\rho$ from the data are well known for various phenomenological models. 
A proper determination depends on the analytical form used to parametrize the nuclear interaction and on the interference with the Coulomb interaction. The quality of the experimental data in the interference region is crucial for this purpose. In the present model there is  analytical connection between the real and imaginary parts, with  control of the fit instabilities, fixing $\rho=\pi/\log(s)$
from dispersion relations. Of course a precise determination of $\rho$ depends on additional terms of a more complete amplitude and on data with good quality in the relevant range.

 In a recent study of the LHC data on pp elastic scattering with independent real and imaginary amplitudes \cite{us}, specific features of the real part, such as the position of the zero and the magnitude and sign of the amplitudes  were investigated, and the parameters were determined with high precision. In the present work we discuss properties of analyticity and crossing symmetry of the amplitudes in the forward regime.   We assume that CM energies 7 and 8 TeV energies are high enough to investigate the scaling property of the amplitudes and introduce a specific model through Eqs.(\ref{amplitude_pomeron_4}), (\ref{tauL}) and (\ref{damping_f1}). We obtain the zero of the real part that influences the form of $d\sigma/dt$. The parameters of $f(\tau')$ lead to predictions for higher energies. Dispersion relations for amplitudes and for derivatives are automatically satisfied. 
 We give predictions for $\sqrt{s}=$ 13 and 14 TeV and also for the cosmic ray energy domain (57 TeV). We also estimate the asymptotic behaviour of the ratio $\sigma_{\rm el}/\sigma$. Precise  measurements expected for high energies $\sqrt{s}=$ 13 and 14 TeV at LHC will allow to test the scaling property of scattering amplitudes. 
  
 \begin{acknowledgments}
The ideas of this work emerged from discussions in the EDS Blois 2017 Conference in Prague, particularly with Profs. A. Martin and O. Nachtmann. The author thanks Profs. Erasmo Ferreira and Takeshi Kodama for useful discussions and for stimulating this work and  careful reading of the manuscript. The author thanks the Instituto de F\'isica of Universidade Federal do Rio de Janeiro for the hospitality. 
The author also thanks the Brazilian agencies CNPq and CAPES for financial support. This work is a part of the project INCT-FNA Proc. No. 464898/2014-5.
\end{acknowledgments}

\end{document}